\documentclass[preprint,5p,times,twocolumn]{elsarticle}

\usepackage{graphicx}
\usepackage{amssymb}

\usepackage{lineno}

\biboptions{authoryear}

\usepackage{multirow}
\usepackage{amssymb}
\usepackage{subfig}
\usepackage{float}
\usepackage{lscape}
\usepackage{amsmath}
\usepackage{array}
\usepackage{textcomp}
\usepackage{multirow}
\usepackage{soul,color}
\usepackage{enumitem}
\usepackage{booktabs}
\usepackage{multirow}
\usepackage{xcolor} %
\usepackage{multirow}
\usepackage{tabularx} 
\usepackage{lscape}
\usepackage{longtable}
\usepackage[mathscr]{euscript}
\usepackage[]{dcolumn}
\usepackage[noabbrev,capitalise]{cleveref}
\usepackage[autostyle]{csquotes}
\usepackage{tabulary}
\usepackage{longtable,array,ragged2e}
\newcolumntype{L}[1]{>{\raggedright\arraybackslash\hspace{0pt}}m{#1}}
\newcolumntype{C}[1]{>{\centering\arraybackslash\hspace{0pt}}p{#1}}
\newcolumntype{R}[1]{>{\raggedleft\\\arraybackslash\hspace{0pt}}m{#1}}
\usepackage{graphicx}
\usepackage{booktabs,dcolumn,caption, makecell}
\newcolumntype{d}[1]{D{.}{.}{#1}} 
\usepackage{bookmark}

\newcommand{\ie}{\textit{i.e.}, }
\newcommand{\eg}{\textit{e.g.}, }

\usepackage{xcolor,colortbl}
\hypersetup{
    colorlinks,
    linkcolor={red!50!black},
    citecolor={blue!50!black},
    urlcolor={blue!50!black}
}
\modulolinenumbers[2]

\journal{-}

\begin{document}

\begin{frontmatter}

\title{How response designs and class proportions affect the accuracy of validation data}

\author[ucl]{Julien Radoux}
\author[csiro]{Fran\c{c}ois Waldner}
\author[ucl]{Patrick Bogaert}

\address[ucl]{Universit\'e catholique de Louvain, Earth and Life Institute - Environment, Croix du Sud, Louvain-la-Neuve, Belgium}
\address[csiro]{CSIRO Agriculture \& Food, Queensland Bioscience Precinct, 306 Carmody Road, St Lucia, 4067 QLD, Australia}

\begin{abstract}
Reference data collected to validate land cover maps are generally considered  free of errors. In practice, however, they contain errors despite all efforts to minimise them. These errors then propagate up to the accuracy assessment stage and impact the validation results. For photo-interpreted reference data, the three most widely studied sources of error are systematic incorrect labelling, vigilance drops, and demographic factors. How internal estimation errors, \ie errors intrinsic to the response design, affect the accuracy of reference data is far less understood. In this paper, we analysed the impact of estimation errors for two types of legends (binary and multi-class) as well as for two common response designs (point-based and partition-based) with a range of sub-sample sizes. We showed that the accuracy of response designs depends on the class proportions within the sampling units, with complex landscapes being more prone to errors. As a result, response designs where the number of sub-samples are fixed are inefficient, and the labels of reference data sets  have inconsistent confidence levels. In order to control estimation errors, to guarantee high accuracy standards of validation data, and to minimise data collection efforts, we proposed to rely on confidence intervals of the photo-interpreted data to define how many sub-samples should be labelled. In practice, sub-samples are iteratively selected and labelled until the estimated class proportions reach the desired level of confidence. As a result, less effort is spent on labelling obvious cases and the spared effort can be allocated to more complex cases, thus leading to an increased reliability of reference data and in turn better accuracy assessment. Across our study site, we demonstrated that such an approach could reduce the labelling effort by 50\% to 75\%, with greater gains in homogeneous landscapes. We  contend that adopting this optimisation approach will not only increase the efficiency of reference data collection but  will also help deliver reliable accuracy estimates to the user community.
\end{abstract}

\end{frontmatter}



\section{Introduction}\label{sec:introduction}

Over the last decades, the number of openly-available geographic data sets has tremendously increased along with their use in policy making, environmental monitoring, hazard prevention and scientific studies. It is of paramount importance that their quality is rigorously evaluated to inform users about their limitations and  to limit contradicting results. Good practices in accuracy assessment include recommendations about i) the sampling design that determine how many sampling units should be collected along with their locations, ii) the response design that defines the protocol for labelling each sampling unit,  and iii) a rigorous estimation of accuracy using specific metrics~\citep{congalton1991review, stehman1998design, stehman1999basic, stehman2001statistical, olofsson2014good}. A statistically rigorous assessment is thus a combination of a probability sampling design, appropriate accuracy estimators, and a response design chosen in accordance with features of the mapping and classification process.\\

Uncertainties linked to the sampling design and variance of the estimators are usually well quantified, but validation methods typically assume that reference data are error-free. In fact, the process of determining the so-called ``ground truth'' is seldom discussed in the literature and is often considered a straightforward --yet costly-- task. Nonetheless, generating authoritative reference data sets remains a major challenge in accuracy assessment and it merits greater consideration in accuracy assessment \citep{Stehman2019111199}. Errors can indeed alter the process of generating reference data and even a small amount of errors can propagate and significantly impact the accuracy assessment~\citep{woodcock2000fuzzy, foody2009impact, foody2013ground}. There is thus a need for new methods that offer better control over the quality of reference data.\\

Practical constraints such as accessibility to the sample locations often affect the implementation of the accuracy assessment. For instance, this prompted~\citet{stehman2001statistical} to propose criteria for quality and statistical rigour while taking, at the same time, practical utility (accessibility and reduced costs) into account. Another alternative is to replace ground observations with photo-interpreted very high-resolution images. Photo-interpretation by a group of experts having regional knowledge is generally seen as the gold standard for reference data collection, especially when dealing with large-area thematic products. Photo-interpretation is not perfect --it typically reaches $80\%$ accuracy-- and it varies considerably among operators \citep[with accuracy levels ranging from $11\%$ up to $100\%$][]{van2014variability}.  Errors ($e$) affecting photo-interpreted samples can be divided in three categories: vigilance, systematic, and estimation errors: \\

\begin{equation}
    e = e_{vigilance} + e_{systematic} + e_{estimation} 
\end{equation}
 
Vigilance errors, \ie loss of performance after performing the same monotonous task over a long period, has been highlighted for a wide range of visual interpretation tasks  ~\citep{parasuraman1986vigilance}. Attitude, either optimistic or pessimistic, may determine how an operator will respond to training for vigilance~\citep{szalma2006training}. Drops of vigilance, which are difficult to predict and manage for an individual interpreter, can be reduced by relying on more than one operator \citep{Pengra2019111261}.\\ 

Systematic errors ($e_{systematic}$) occur when a photo-interpreter is incorrectly reading the image. Reading of images and maps belongs to cartographic and visual literacy~\citep{svatovnova2017reading}. Visual literacy skills change over time and can be improved with the development of geospatial thinking. Image interpretation is a process that combines perception and cognition, both of which tend to facilitate identification (the cognitive task of identifying a pattern) and signification (the assignment of a meaning to a particular pattern~\citep{olson1960elements, colwell1965extraction}). The types of insight derived from imagery are strongly influenced by the interpreter's expertise. Experts bring specialised knowledge, highly attuned perceptual skills and flexible reasoning abilities that novices lack~\citep{klein1993perceptual}. There is however no strong relationship between the fieldwork experience of operators and their photo-interpretation accuracy. This might be explained by the dissimilarities between, on one side, air- and spaceborne images and, on the other side, panoramic images in at least three important aspects: i) the portrayal of features from a downwards --often unfamiliar-- perspective, ii) the use of wavelengths outside the visual portion of the spectrum, and iii) the depiction of the Earth's surface at unfamiliar scales and resolutions~\citep{lillesand2008digital}. The most capable interpreters have keen powers of observation, coupled with imagination and a great deal of patience \citep{lillesand2008digital}. Another individual factor potentially influencing image interpretation accuracy is search strategy. Compared to random search, training in systematic inspection caused higher performance~\citep{wang1997training}. \citet{maruff1999behavioral} have suggested that behavioural goals constrain the selection of visual information more than the physical characteristics of the information. This suggests that photo-interpreters with a search strategy based on previous experiences would be more successful at extracting relevant information than someone randomly searching for this information. Geographers would therefore be more successful than non-geographers during a single categorisation round of aerial photos~\citep{lloyd2002visual}. Accordingly, crowdsourcing (\ie when photo-interpreters are replaced by volunteers) is particularly prone to errors as it is open to anyone, regardless of the  level of expertise  of the volunteers. Systematic errors can thus largely be avoided by providing training to photo-interpreters, selecting operators with local knowledge, and relying on multiple contributors \citep{Pengra2019111261}.  \\

Estimation errors ($e_{estimation} $) arise when the class proportions within sampling units are imprecise even when all labels related to the sampling units are correct. These errors stem from three main factors: the number of sub-samples to label per sampling unit, the landscape structure, and  the choice of the legend. Imprecise estimates of the proportion of different classifiers for mixed or transitional classes reportedly account for the majority of disagreements among photo-interpreters~\citep{powell2004sources}. It has also been shown that the accuracy of the labelling as well as the accuracy of the image-based classification generally decrease when the sub-pixel heterogeneity increases \citep{tran2014}.  Contrary to the systematic and vigilance errors, there is currently no mechanism to control estimation errors. That is when best practices in quality control are implemented, \ie $e_{systematic} = 0$ and $e_{vigilance} = 0$, errors in the photo-interpreted labels remain and are solely due to estimation errors. If left unchecked, estimation errors can thus bias reference data as they are intrinsically linked to the complexity of the sub-pixel landscape structure. We thus proposed that estimation errors need to be managed in the response design.\\

With the objective of improving the accuracy of photo-interpreted reference data, we sought to untangle the intricate relationships between legends, response designs and landscape fragmentation with regards to the estimation errors. Our specific goals were i) to estimate the impact of imprecise estimation of land cover proportions on the accuracy of reference data and ii) to propose a response design that optimises the labelling effort. We particularly focused on two aspects of response designs: their structure (point-based \textit{vs.} partition-based designs) and their associated labelling effort (the number of sub-samples to be labelled per sampling unit). These were studied for both binary an multi-class legends. Our main contributions can be summarised as follows:
\begin{itemize}
    \item We provide an in-depth review of the different types of response designs and their applications;
	\item We analyse the performance of response designs for different types of legends;
    \item We generalise case-specific results using landscape indices;
    \item We optimise the sampling effort with adaptive response designs that leverage the confidence intervals of the estimated proportions.
\end{itemize}
\noindent To investigate estimation errors independently of other sources of error, we simulated different realisations of response designs and legends based on a 2-m land cover map of Wallonia, Belgium, that we considered as ground-truth. To isolate the effect of estimation errors, we  assumed a perfect quality control system throughout the paper, \ie there were no vigilance or systematic errors. 

\section{Background}
\subsection{Types of classification system}

There is no one ideal classification system and it is unlikely that one could be developed \citep{anderson1976land}. A large number of classification systems have therefore been developed depending on the purpose of the map and the scale of the analysis. \citet{congalton2014global} suggest that all classes of the legend should be clear determined at the beginning of the mapping projects, which is not always the case \citep{grekousis2015overview}.  Those classification systems can be divided into three categories: semantic legends, aggregative legends and continuous fields. \\

A semantic legend provides a formal description of the classes, including properties and relationships with their sub-parts. They are usually preferred to describe spatial entities at large scales like \eg trees or buildings. Spatial entities unambiguously described by a semantic legend can in turn be used as diagnostic criteria~\citep[also referred to as classifiers in the context of classification systems;][]{di2005land} to build aggregative legends. In some cases, a semantic legend can be used to describe spatial regions instead of spatial entities \citep{radoux2017good}. In this case, the semantic refers to a specific meaning that encompasses a large set of properties and relationships, \eg a city (land use type) or a savanna (ecosystem type). To avoid ambiguity, all classes should be precisely described by an ontology that includes a representation, a formal naming, and a definition of their properties and relations \citep{arvor2019}.\\

When the spatial resolution of the image becomes coarser than the size of the spatial entities, pixels can either represent a continuous field with the proportions of all (or a subset of) the classifiers \citep{fernandes2004approaches}, or can be described by an aggregative categorical legend. For the latter, the proportion of spatial entities is computed and decision rules are applied to define the classes at the coarser resolution \citep{petit2001integration}. These decision rules of aggregative legends are either based on a majority rule or on fixed threshold values for class proportions.  \\

Majority-based legends are the most prevalent type of aggregative legends in land cover mapping. For instance, all the first global land cover maps used of IGBP (International Geosphere-Biosphere Programme) legend \citep{grekousis2015overview}. It is often implicitly assumed (without being always clearly stated) that labels correspond to the class of the prevalent spatial entity within each mapping unit. The boundary between a pure semantic legend and a majority-based legend is therefore fuzzy. The drawbacks are that (i) majority is undefined when multiple classes are equally dominant,  and (ii) no information about the actual class proportions within the mapping units is conveyed to users. As an example, in the case of a legend with ten classes, a majority label could be assigned to a class covering between 10\% (all classes present in the same proportion) and 100\% (``pure'' class) of the area of the mapping unit.\\

Threshold-based legends are also a widely used type of aggregative legends. They rely on a set of rules to partition the feature space of the classifiers' proportions. Those rules introduce sharp boundaries to the continuous field of the classifiers' proportions in order to obtain a limited number of categorical classes. The number of those classes usually exceeds the number of classifiers in the mapping area. Threshold-based legends should be defined with consistent classification systems such as the Land Cover Classification System \citep[LCCS;][]{di2005land}. The LCCS guarantees no overlap between classes and a full description of the possible combinations of classifiers and is considered by \citet{grekousis2015overview} as the only universally applicable classification system. LCCS is used for coarse to medium resolution global land cover maps \citep{bartholome2005glc2000, bontemps2013consistent}, but also for high resolution object-based classification \citep{radoux2017good}. They provide a larger thematic precision (more classes from the same set of classifiers) than majority-based legend, but they require to define (often arbitrary) thresholds and conditional statements to draw the boundaries between the classes. This may lead to difficult naming conventions when the feature space of classifiers is large and the rules become complex. \\

Binary legends are a particular case of aggregative legends that indicates whether a given classifier is present or absent inside a given pixel. Several examples include global crop/non cop \citep{waldner2016unified}, water/non water \citep{lamarche2017compilation} or forest/non forest \citep{SHIMADA201413} maps. The labelling is then defined by a proportion threshold. Most of the time, this threshold is set to 50\%, which is then nearly equivalent to a majority-based legend for two classes, but can be unambiguously defined at the proportion of 50\% (unlike the majority). There are however examples where the threshold is not 50\%, \eg 10\% for a forest/non forest map in arid regions \citep{bastin2017extent}.

\subsection{Structures of response designs}

\begin{figure}[h!!!]
\centering
\includegraphics[width=0.8\linewidth]{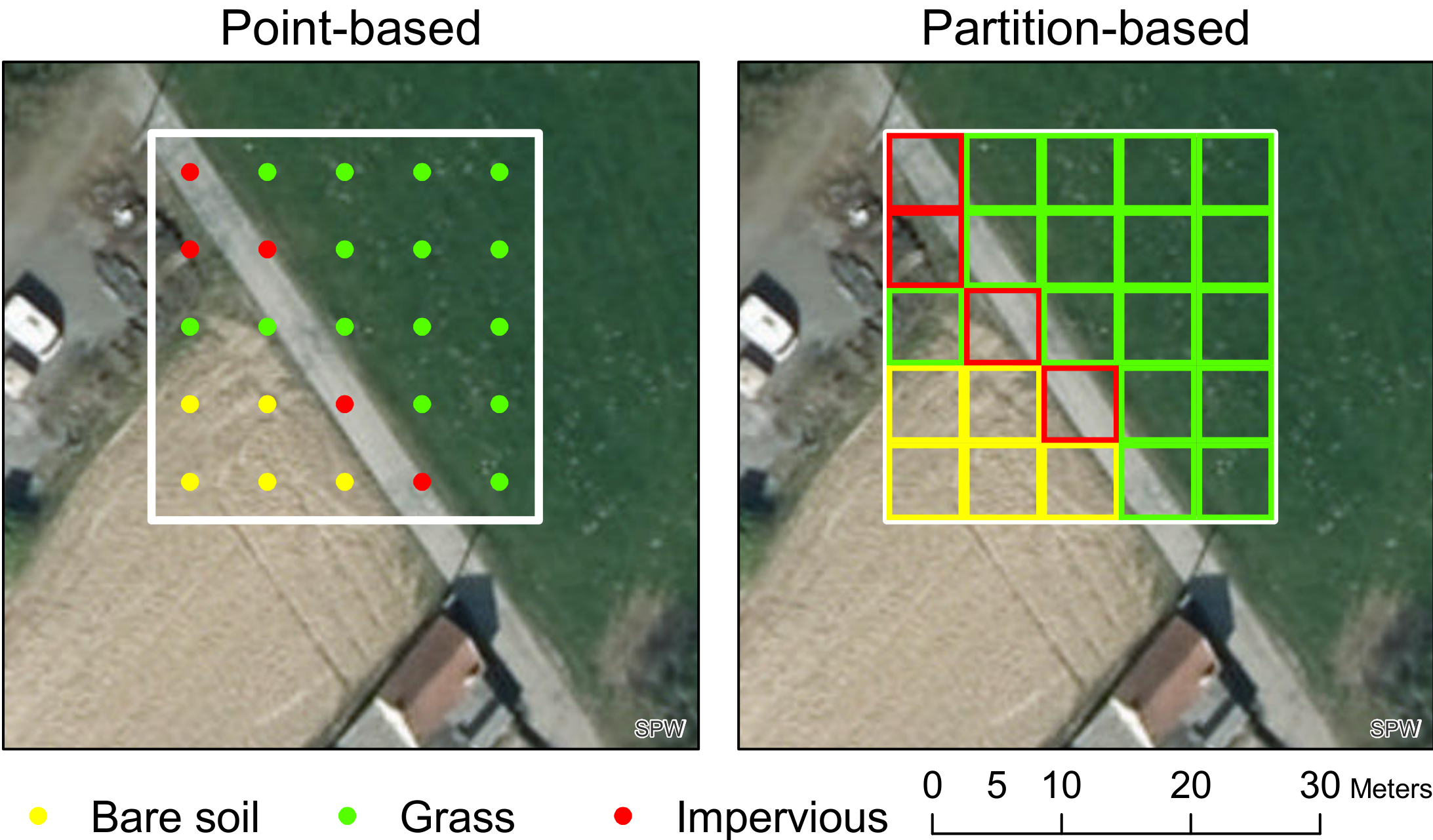} \caption{\label{fig:sampling_type} Two types of response designs compared in this study: point-based and partition-based designs. Grassland are in green, impervious surfaces are in red and cropland are in yellow. Both methods would agree on a majority of grassland, as well as the direct assignment of the majority class without looking at sub-samples.}
\end{figure}

The values of the accuracy parameters are strongly affected by the protocols implemented for the response design. This includes the choice of spatial units and how within-unit homogeneity is addressed when assigning class labels \citep{stehman2011pixels}. Most of the time, the sampling unit is the elementary unit that is labelled according to the legend of the map. In this study, the sampling unit is a pixel and there is an agreement when the map classification and the response design converge to the same label. The studied response design is therefore crisp, as opposed to alternative response designs where the proportions of spatial entities are used to validate categorical variables, \eg  fuzzy validation. Best practices in accuracy assessment suggest that photo-interpretation should rely on images of finer resolution than the map being validated. With finer resolution data, the sampling unit could appear as heterogeneous. Three protocols could then be used to assign a reference value to those sampling units: 
\begin{description}
\item[Direct assignment:] a single label is directly assigned to the sampling unit;  this is the most common response design~\citep[see][]{perger2014cropland};
\item[Point-based sub-sampling:] the sampling unit is sub-sampled by a set of points that are individually labelled by the interpreter~\citep[see][]{bey2016collect, bastin2017extent} before automated aggregation with decision rules; 
\item[Partition-based sub-sampling:] the sampling unit is partitioned into sub-parts that are labelled individually by the interpreter~\citep[see][]{bayas2017global, waldner2019conflation} before automated aggregation with decision rules.
\end{description}
\noindent Direct assignment is the fastest method because a single class is assigned for each sampling unit by looking at it as a whole. However, (i) it is strongly dependent on the operator skills and its level of concentration, (ii) it is poorly inter-operable because there is no information about classifiers proportion that would help to apply other labelling rules, and (iii) the confidence of the labelling must be provided by the operator. In this study, we focused on the two designs that involve sub-samples, namely point-based and partition-based  designs (Fig. \ref{fig:sampling_type}) because our primary assumption was that a large part of the labelling errors could be quantified from the selected response design, \ie independently from the operators.  \\

\subsubsection{Point-based response designs}

In point-based response designs, photo-interpreters label a set of points within every sampling unit. The final class is then assigned based on the proportions of the number of points for each observed category. By definition, points are dimensionless but, in practice, the photo-interpretation is limited by the spatial resolution of the reference image. Nevertheless, even if the vicinity of points provides contextual information as a part of the photo-interpretation process, the label is defined at the precise location of the point. \\

Randomly selecting a set of points inside each sampling unit inherits the same properties as for the sampling designs of a map as a whole. Systematic sampling is therefore often the most efficient \citep{stehman1992comparison}. However, if strong periodicity in the spatial pattern of the landscape is suspected, systematic sampling should be avoided unless sufficient information is available to avoid the phasing between spatial pattern and sampling interval \citep{matern2013spatial}. \\

\subsubsection{Partition-based response designs}

Contrary to their point-based counterparts, partition-based response designs provide exhaustive coverage of the sampling unit (Fig. \ref{fig:sampling_type}). The counterpart is that it implies a discretisation of the landscape, which could result in inaccurate labels. In practice, photo-interpreters are tasked with estimating the proportion of each class within each sub-sample, based on their entire content. The final label is automatically attributed from the estimated class proportions following a set of rules that are specific to the legend. These rules are applied in a two-step process: the first step is performed by a photo-interpreter who assigns a label to each sub-part, and the second step consists in aggregating the labels of the sub-parts to attribute a final label, which can be automated.  While square sub-parts are the most widespread type of partition~\citep{bayas2017global}, irregular polygons can also be used~\citep{waldner2019conflation}. In the latter case, accurate delineation of the polygons plays a major role in the reliability of the response design. \\

In the case of binary classification, there are two approaches to define the sampling unit labels. The first approach, hereafter referred to as Threshold-Then-Majority (TTM), applies the labelling rule for each sub-sample level, then determines the majority label amongst sub-samples of the whole sampling unit. The second protocol, hereafter referred to as Majority-Then-Threshold (MTT), starts by determining the majority class inside the sub-samples, then applies the labelling rule with threshold at the level of the sampling unit. These methods are identical when the binary threshold is equal to $50\%$ but differ otherwise. These two frameworks could be applied to LCCS-like multi-class legends as well. In the case of majority class legend, they both simplify into a two-stages application of the majority rule. TTM is expected to work best in fragmented landscapes and MTT for large homogeneous patches. 

\section{Data}\label{sec:data}

\subsection{Generating sampling units and sub-samples from a reference land cover map}
  
\begin{figure}[htb!]
\centering
\includegraphics[width=0.48\textwidth]{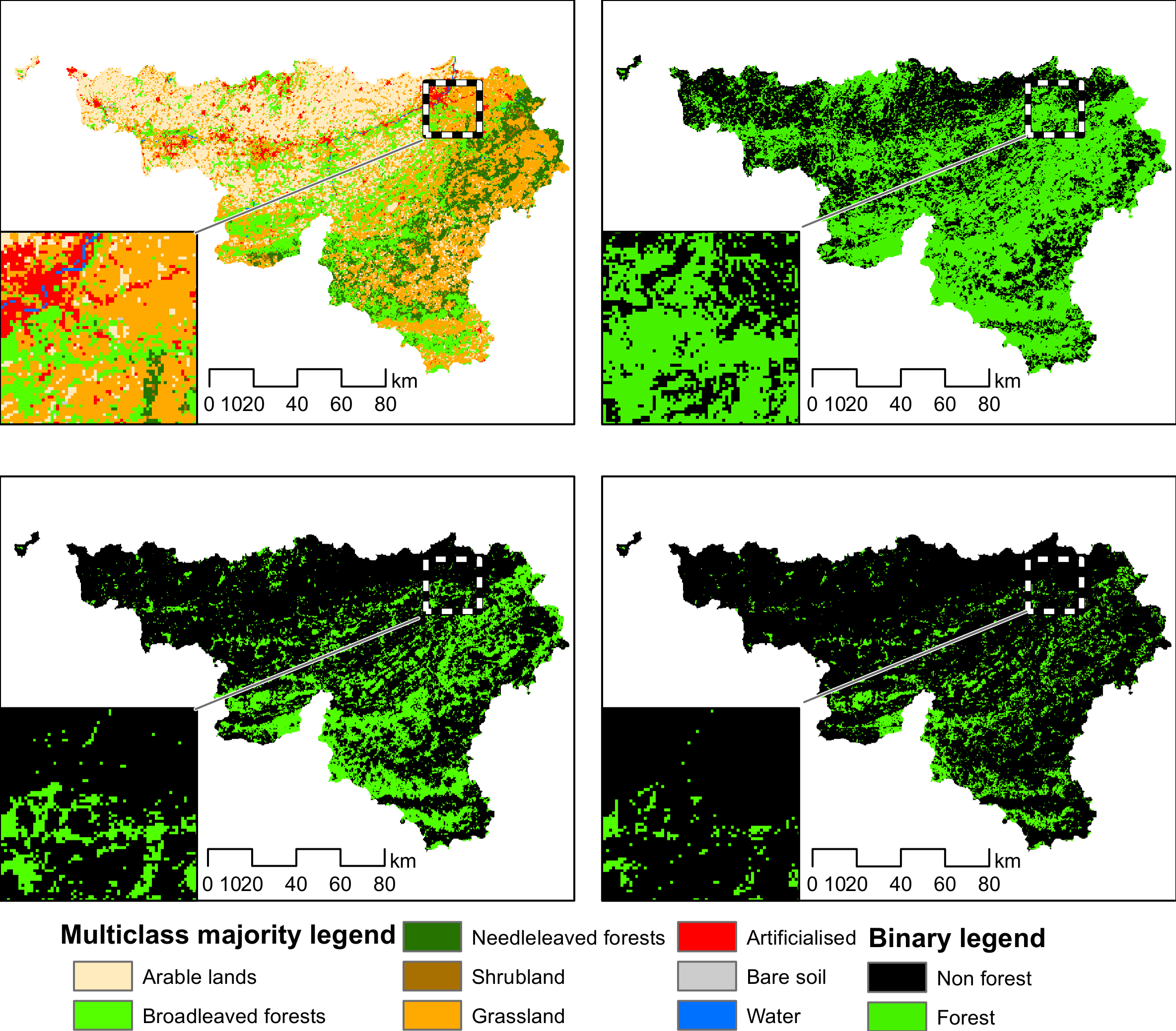}\caption{\label{fig:legend} Maps of the study area based on majority-based legend and the different threshold-based binary legend for the forest classes: 10\% for top right, 50 \% for bottom left and 75\% for bottom right.  }
\end{figure}

To obtain an error-free reference data set across resolutions, we considered a 2-m land cover map  as ground truth~\citep[Fig.~\ref{fig:legend};][]{radoux2019ecotopes}. The original map covers the Walloon Region, Belgium, and includes ten land cover classes. The two marginal grassland classes were merged with the agricultural grassland because of their scarcity. Using a 2-m resolution map provides unambiguous labelling of classifiers because the size of the pixels is inferior to the size of standard objects in this landscape. In case of mixed pixels located at the boundary of two spatial objects, the label was chosen based on the pixel centroid. \\

The overall accuracy of the land cover map was estimated based on 1000 points selected by stratified probabilistic sampling. Sample units were photo-interpreted on 25 cm resolution orthophotos by two trained operators. Sampling units with conflicting or uncertain labelling were verified \textit{in situ} with sub-metric positioning. The estimated overall accuracy of the 2-m land cover map was 85\%, this including both thematic and geolocation errors, and it reaches 93\% when geolocation errors of less than 5 meters are tolerated. Hence the map was expected to exhibit realistic land cover patterns. Henceforth, we considered the 2-m land cover map as an exact reference, \ie spatially precise and thematically accurate.\\ 

\begin{figure*}[h!!]
    \centering
    \includegraphics[width=0.99\textwidth]{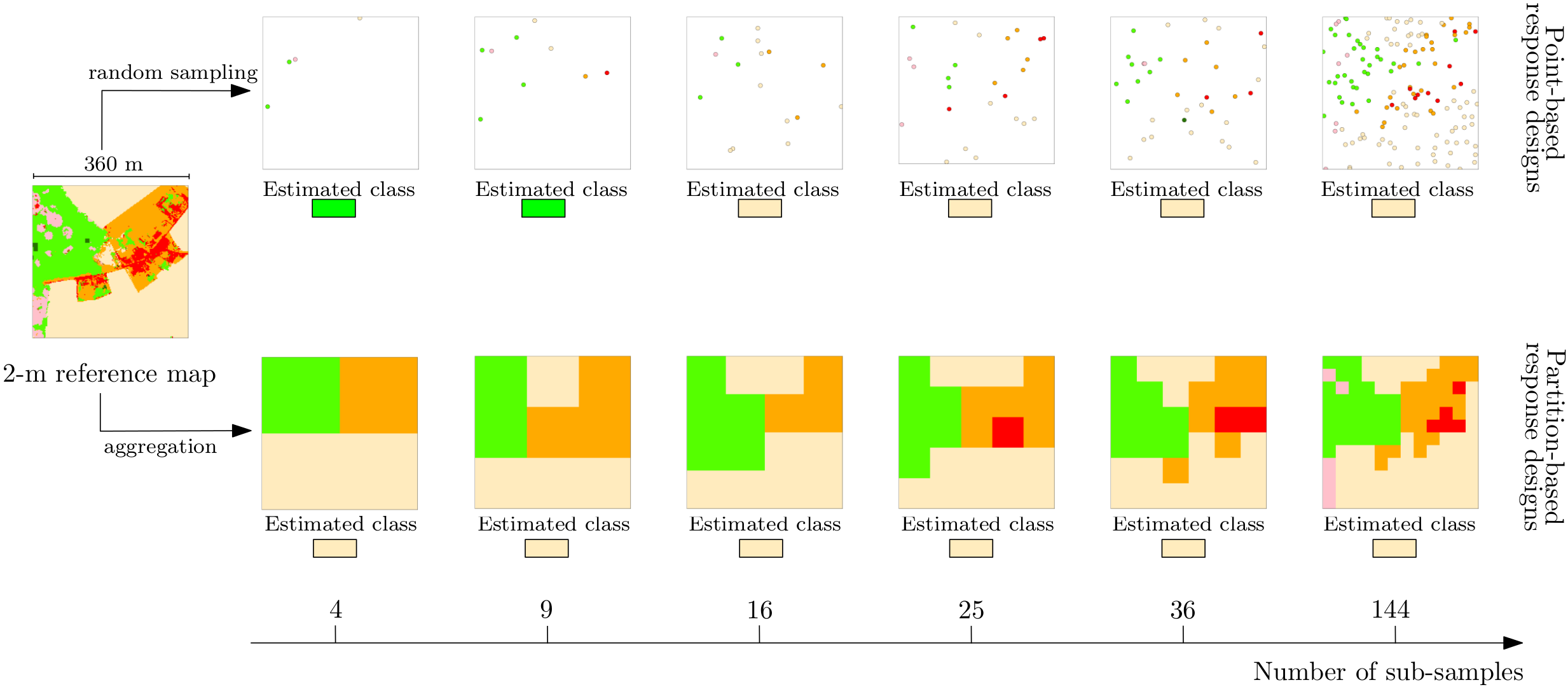}
    \caption{Procedure to generate sampling units and sub-samples from the 2-m reference map.}
    \label{fig:resampling}
\end{figure*}

In this experiment, pixels in the map at 360-m resolution corresponded to sampling units and their labels were considered as the target values for the different response designs. The high resolution data set was aggregated at the resolution of 360 m using the different labelling rules based on the pixel counts (with a number of 32400 2-m pixels per 360-m cell; Fig.~\ref{fig:resampling}). While 360 m does not correspond to the spatial resolution of any current satellites, it is similar to medium-resolution satellite such as PROBA-V, MODIS, Sentinel-3 or VIIRS and it has the advantage of being factorisable by a diversity of integers (2, 3, 4, 5, 6, 9, 12, 15, 20, \ldots), thereby allowing partitions of a large variety of size of squares.   

\subsection{Types of legends}
\label{sec:legend}

Two types of aggregative legends were compared in this study: a multi-class majority-based legend and a set of three threshold-based binary legends (Fig.~\ref{fig:legend}). With the majority-based legend, the label of the most frequent class within the mapping unit was selected. There are eight classes in the 2-m map so there are also eight classes at the aggregated level. In the cases of binary legends, the map represents the presence or absence of a specific class. When pixels are larger or similar in size to the spatial objects or the spatial regions of interest, an arbitrary threshold on the proportion of the class becomes necessary to handle mixed pixels. In this study, we propose to look at maps of the forest class (broadleaved and needleleaved forests together). Three different thresholds of crown cover have been chosen, namely 10\%~\citep[FAO's forest definition][]{FAO2010FRA}, 50\%  (the most commonly-used threshold) and 75\% (threshold used for closed canopy forests). 


\section{Methods}\label{sec:method}

We sought to answer the following questions:
\begin{enumerate}
\item What is the accuracy of point-based and partition-based response designs for different number of sub-samples in a realistic case study?
\item How can the accuracy of response designs be predicted based on landscape structure indices?
\item How to optimise the number of sub-samples per sampling unit?
\end{enumerate}
\noindent We addressed these questions in three successive steps for the four legends described in section \ref{sec:legend}. First, the accuracy  of the various response designs was compared with simulated sampling across the study site  (section~\ref{sec:accuracy}). Second, we generalised the relationship between error rates and the  landscape structure  underlying the sampling units (section~\ref{sec:frag}). We finally proposed a method that iteratively adds sub-samples to label until the estimated class proportions driving the labelling process reach a the desired confidence level  (section~\ref{sec:opt}). \\

This optimisation methods is formulated for point-based designs only, as theoretical confidence intervals are not available for partition-based designs and  labels cannot be reused when the number of partition is increased. In fact, optimising the partition-based designs depends on the ability of the operator to decide the appropriate number of sub-samples. As we assumed perfect operators throughout this paper, the question of optimising partition-based designs falls beyond the scope. Nonetheless, the impact of photo-interpretation errors on the response design is  discussed in section~\ref{sec:discussion}. \\

\subsection{Accuracy of point-based and partition-based response designs}\label{sec:accuracy}

Our approach to empirically quantify the accuracy of response designs was based on a Monte Carlo framework. For every sampling unit, we repeatedly estimated the class proportions of ground truth for a range of sub-sampling efforts. The labels of the randomly simulated response designs were then assigned using the decision rules of the four legends. The same decision rules were applied on the true proportions (\ie computed from the 2-m reference raster) to derive the true label. For each iteration, the error rate was computed by dividing the number of disagreements between the true and the simulated labels of the sampling units, with

\begin{equation}
    \text{Error rate} = \frac{\text{Number of erroneous labels}}{\text{Number of sampling units}}
\end{equation}

For the point-based response designs, the sub-sample selection was repeated 36 times. Sub-samples were selected by simple probabilistic sampling of the 2-m pixels located inside each sampling unit. \\

For partition-based response designs, the 36 realisations were generated by shifting the origin of the grid by six multiples of eleven pixels (22, 33, 44, 55, 66, 77) in both the $X$ and $Y$ directions (resulting sampling units that are not completely inside the study area are discarded). Spatial resampling of the 2-m reference map was performed at intermediate spatial resolutions of 180, 120, 90, 72, 60, 36 and 30 m that respectively correspond to a partitioning in 4, 9, 16, 25, 36, 100 and 144 squares. In the TTM case, the proportion of the forest class was computed within every intermediate resolution pixel, which were then labelled as forest or non-forest according to the threshold value. Those pixels were then resampled at the spatial resolution of 360 m with a majority rule to select the final label. In the MTT case, the forest label was assigned to each sub-sample where forest was majority. The proportion of forest pixels was then computed for each 360 m pixel and the final forest/non-forest label was assigned based on the selected threshold. For the majority legend, the majority class was first identified for each sub-sample and the majority of the sub-sample labels was finally assigned to the sampling unit. 

\subsection{Impact of landscape fragmentation}\label{sec:frag}

In this section, we sought to evidence the link between landscape structure and response design in order to predict the response design accuracy in other landscapes where \emph{prior} structure knowledge is available. We therefore selected two landscape metrics --one per legend type-- that can be easily computed for any areal sampling unit and any scale.\\

For binary legends, we characterised landscapes by reporting the purity ($\pi$) of the forest class within each sampling unit, with
\begin{equation}
\pi = \frac{S_{\text{forest}}}{S_{\text{total}}}
\end{equation}
\noindent where $ S_{\text{forest}}$ is the area of forest (more precisely in this case, tree crown cover) inside the sampling unit, and  $S_{\text{total}}$ is the area of the sampling unit. \\

For multi-class legends, we opted for the Equivalent Reference Probability~\citep[$\epsilon$;][]{bogaert2016information}. Rooted in information theory,  $\epsilon$ is particularly interesting because it accounts for the full set of probabilities and remains consistent with the maximum probability, unlike entropy. Given  $\mathbf{p} = (p_1, ..., p_k)$ the vector of the class proportions in the landscape, $k$ the number of classes and $i^*$ the index of the dominant class, the Equivalent Reference Probability $\epsilon$ is

\begin{equation}
\label{probequiv}
\epsilon= \frac{\exp \left(E[D(i\vert\vert i^*)]\right)}{\exp\left(E[D(i\vert\vert i^*)]\right)+k-1}
\end{equation}

\noindent where  $E[D(i\vert\vert i^*)]$ is the expected difference of information: 

\begin{equation}
\label{expected gain}
E[D(i\vert\vert i^*)]  = \ln p_{i^*}-\frac{1}{1-p_{i^*}}\sum_{i\backslash i^*}p_i\ln p_i
\end{equation}

\noindent with $p_{i^*}$, the proportion of the majority class. Class purity and $\epsilon$ were computed for each sampling unit based on the true proportions.\\ 

Average error rates of the response designs were estimated for the full range of possible $\pi$ and $\epsilon$ values with a step of $0.05$. For visualisation purposes, the error rate were smoothed by fitting local regressions~\citep[LOESS;][]{cleveland1992local}. 

\subsection{Local optimisation of the number of sub-samples}\label{sec:opt}

When collecting validation data, the structure of the landscapes covered by the sample units is generally unknown, so that the optimal number of sub-sampling units cannot be estimated \textit{a priori} from relationships between accuracy and landscape fragmentation. However, the interactivity of Web 2.0 validation platforms allows us to compute on-the-fly the proportion of each class according to the available sub-samples. This third part of the study aimed to optimise the number of sub-samples needed for reaching a certain level of accuracy, resulting in an optimal response design that minimises cost and/or time constraints. Therefore, we propose to define an optimal number of sub-samples for each sampling unit based on the confidence intervals of the estimated sub-sample class proportions. Here, $\alpha$ was set to 0.001 to illustrate the stringent requirements of building authoritative reference data sets, and to 0.1 to illustrate the required effort for collecting reference data under constrained conditions.\\ 

In practice, the local optimisation process consisted in randomly selecting an initial set of nine sub-samples and assessing the corresponding confidence level. Sub-samples were then added one at a time until the confidence on the estimated proportions is larger than the desired confidence. For the binary legends, the confidence interval (for a given confidence level) around the estimated proportion must not include the threshold value that divides the study area in the two binary classes. For the multi-class majority legend, the confidence interval around the estimated proportions of the majority class must not include the estimated proportion of the second most frequent class. \\

For binary legends, a given sampling unit is correctly labelled if the estimated proportion is on the same side of the threshold value than the true proportion. In practice, the proportion of the sampled area is unknown. However, the probability of assigning the correct label can be estimated based on the estimated value of the binomial distribution. Because of the small number of points, the Clopper-Pearson exact confidence interval (CI) was used instead of the Normal approximation~\citep{clopper1934use}, with

\begin{align}
\label{eq:Clopper-Pearson_lb}
\text{CI}_{LB} &= 1 - \text{BetaInv}(\frac{\alpha}{2},~n-m,~m+1) \\
\text{CI}_{UB} &= 1 - \text{BetaInv}(1-\frac{\alpha}{2},~n-m+1,~m) 
\end{align}

\noindent where $\text{CI}_{LB}$ is the confidence interval lower bound, $\text{CI}_{UB}$ is the confidence interval upper bound, $n$ is the number of sub-sampling units, $m$ is the number of points belonging to the majority class, $\alpha$ is the percent chance of making a Type I error, and $1-\alpha$ is the confidence level. \\

Exact confidence intervals are not available for multinomial cases. Following \citet{goodman1965simultaneous}, simultaneous confidence interval estimates were selected because preliminary tests revealed that, in a binomial case, it provides a closer match to the Clopper-Pearson interval than other alternatives. For a multinomial distribution $\mathbf{p}$, Goodman's simultaneous confidence interval for the $i^{th}$ class is given by

\begin{equation}
\label{eq:goodman}
\text{CI}_{i} = \frac{b + 2p_i \pm \sqrt{b[b+4p_i(n-p_i)/n])}}{2(n+b)}, \ i = 1,\ldots, k 
\end{equation}
\noindent where  $b = \chi^2_{1-\alpha/k}(1)$, the $1-\alpha/k$ quantile of the chi-square distribution with one degree of freedom. \\

In some cases, \eg where the observed class proportion is equal to the arbitrary threshold in a binary classification or when several classes have the same proportion in the case of majority rule, the number of points to meet the required confidence could grow infinitely. Therefore, the maximum number of points was arbitrarily set to 144. This process was repeated 25 times in order to compare the theoretical confidence levels with the observed accuracy and to estimate the average number of sub-samples needed for each sampled area.

\begin{figure*}[h!!!]
\centering
\includegraphics[width=0.7\linewidth]{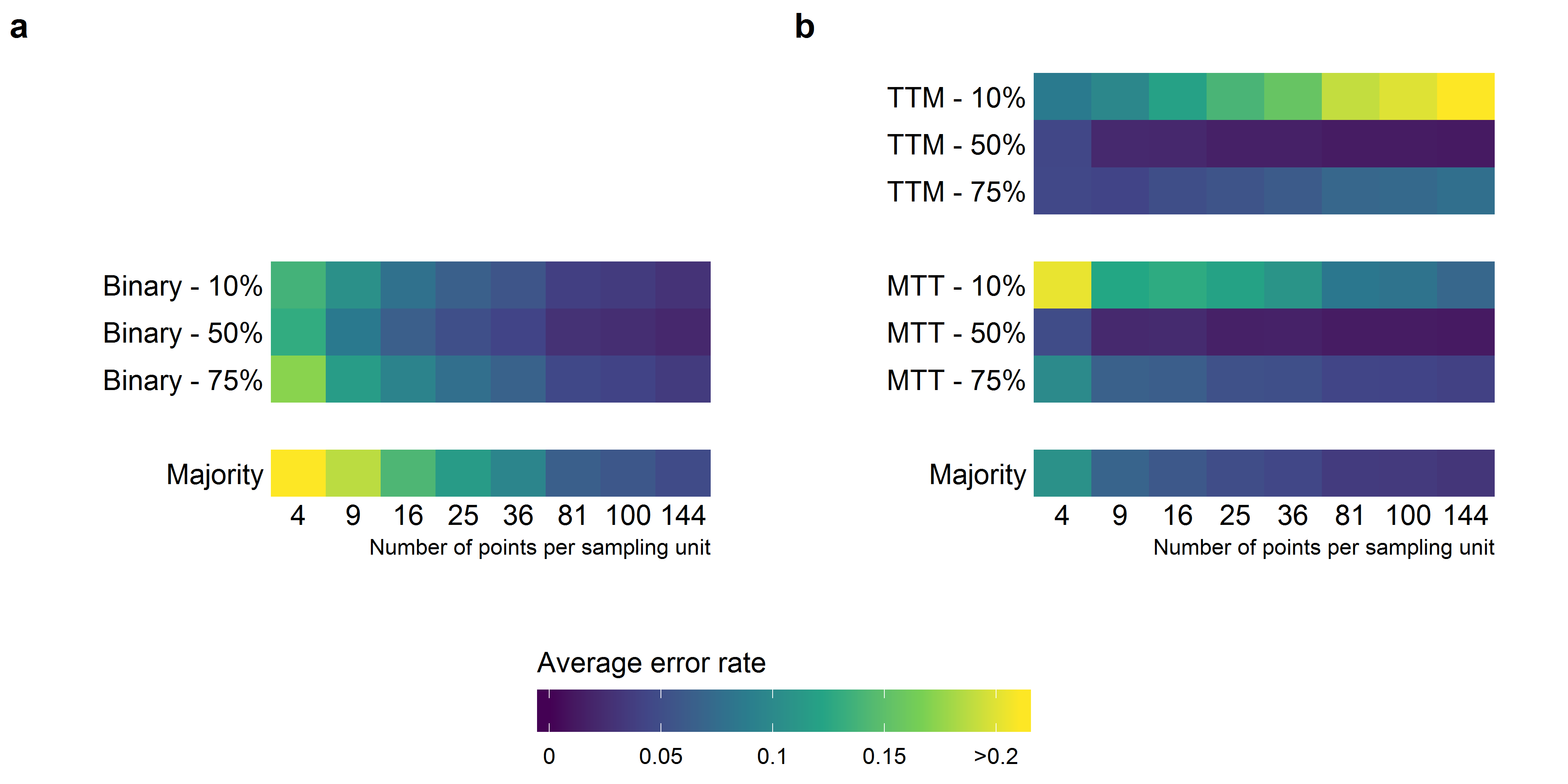}
\caption{\label{fig:error_rate_nsamples} Average error rate in reference labels across the study area for (a) point-based  and (b) partition-based  designs.}
\end{figure*}

\section{Results}

\subsection{Impact of response design and sampling effort on accuracy of the labels} \label{sec:caseStudy}

Overall, results highlight the relatively large uncertainty linked with the response designs for all types of legends in the study area. In addition, the average error rate is not only linked with the sampling effort, but also depends on the combination of the legend and the type of response design (Fig.~\ref{fig:error_rate_nsamples}). \\

For any sub-sample size, the most reliable labels are obtained for the binary legend with a threshold at $50\%$ for both partition-based and point-based response designs. For the other legends, the ranking of the ease of validation differs across response designs. For instance, the second most consistent labelling is obtained for the majority legend with a partition-based design, while the binary legend with $10\%$ threshold ranks second for point-based validation. For the same legend, the partition-based response design performs poorly, with an error rate of $12\%$ for $25$ sub-samples. \\

The average error rates of point-based designs markedly decreases between 4 and 100 sub-samples (Fig. \ref{fig:error_rate_nsamples}a). This trend is observed across the four legends. With only four points, error rates are $>15\%$.  The error rates then drop to $<2.5\%$ for sub-sample sizes larger than $100$ in the case of threshold-based legends. The decreasing error rate with respect to the sampling effort is also observed for the majority legend, but the improvement is smaller (still $6\%$ error with $100$ sub-samples). In comparison, the other binary legends provide more correct labels ($4\%$ error with $100$ sub-samples for the $75\%$ binary legend), with the most accurate labelling obtained from the binary $50\%$ legend ($3.5\%$ error).\\

The two types of partition-based response designs exhibit an opposite behaviour for the binary thresholds of $10\%$ and $75\%$. In those two cases, the error rates of a perfect operator increase in TTM (but decrease in MTT) for increasing numbers of sub-samples. The binary legend at $50\%$ yielded similar results for MTT and TTM, with slightly better results from the TTM approach. It reaches $98.4\%$ accuracy with $25$ sub-samples. The majority-based legend fails to generate labels with less than $3\%$ error when using less than 144 sub-samples, and achieves less than $5\%$ errors starting from 25 sub-samples (Fig. \ref{fig:error_rate_nsamples}b).\\

The results of the case study show that the most efficient response design depends on the legend of the map. Given the spatial resolution of the sampling units and the relatively fragmented landscape of the study area, the partition-based response design outperformed point-based response design for the majority legend. With the latter, $25$ sub-samples were necessary to achieve $95\%$ of accuracy. The validation effort required for binary legends depended on the threshold value. The least effort was required with a threshold of $50\%$ and a partition-based model ($95.2\%$ with only 4 sub-samples). On the contrary, point-based response design outperformed partition-based response designs for the $10\%$ threshold. This legend was the most difficult to validate in the study area --36 sub-samples were needed to reach at least $95\%$ accuracy.   

\subsection{Relationship between sampling unit heterogeneity and accuracy}

\begin{figure*}[htb!]
\centering
\includegraphics[width=0.85\linewidth]{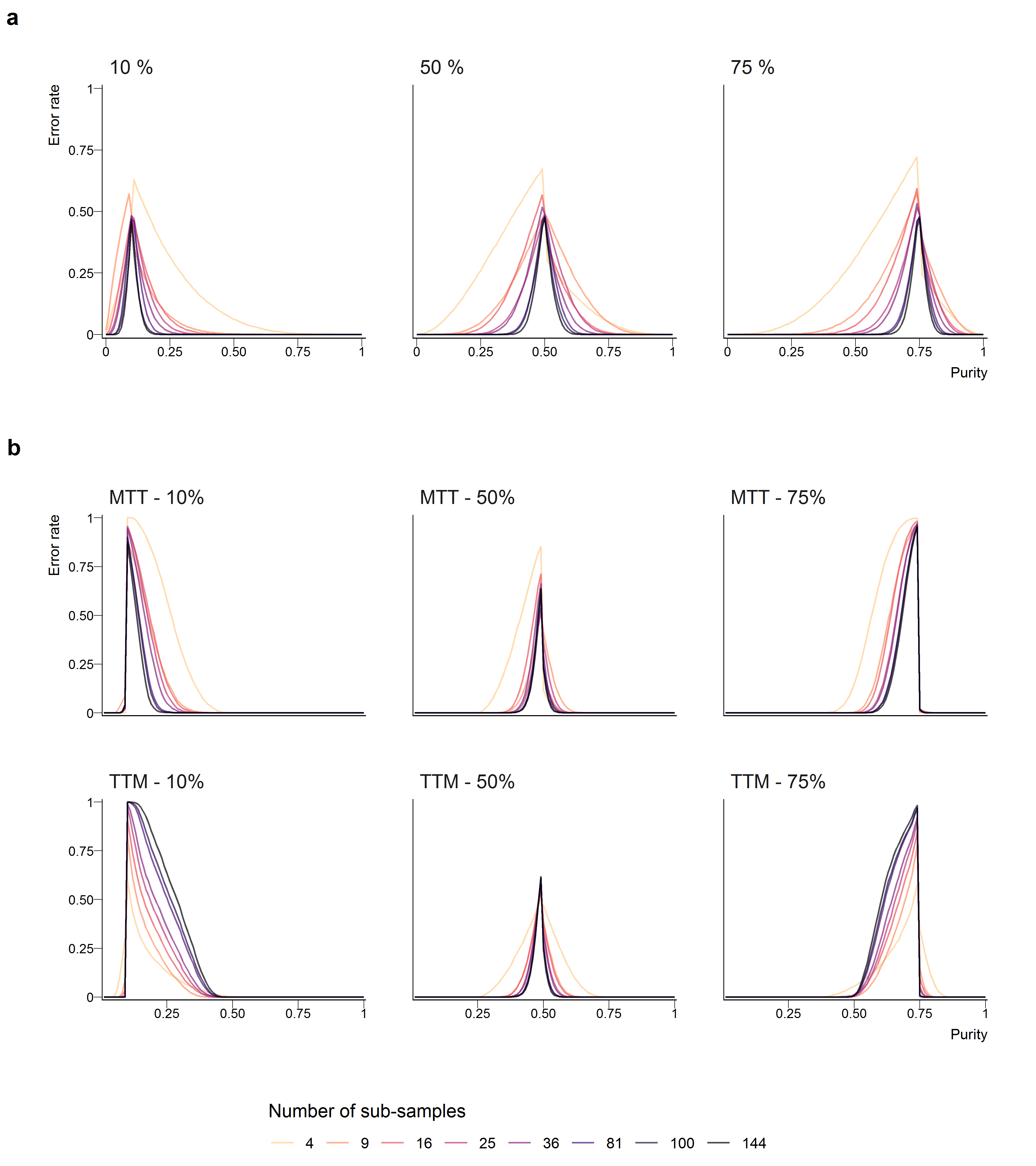}
\caption{\label{fig:error_rate_nsamples_bin} Error rates in reference labels as a function of landscape fragmentation  for binary point-based (a) and partition-based (b) designs.}
\end{figure*}

\begin{figure*}[htb!]
\centering
\includegraphics[width=0.75\linewidth]{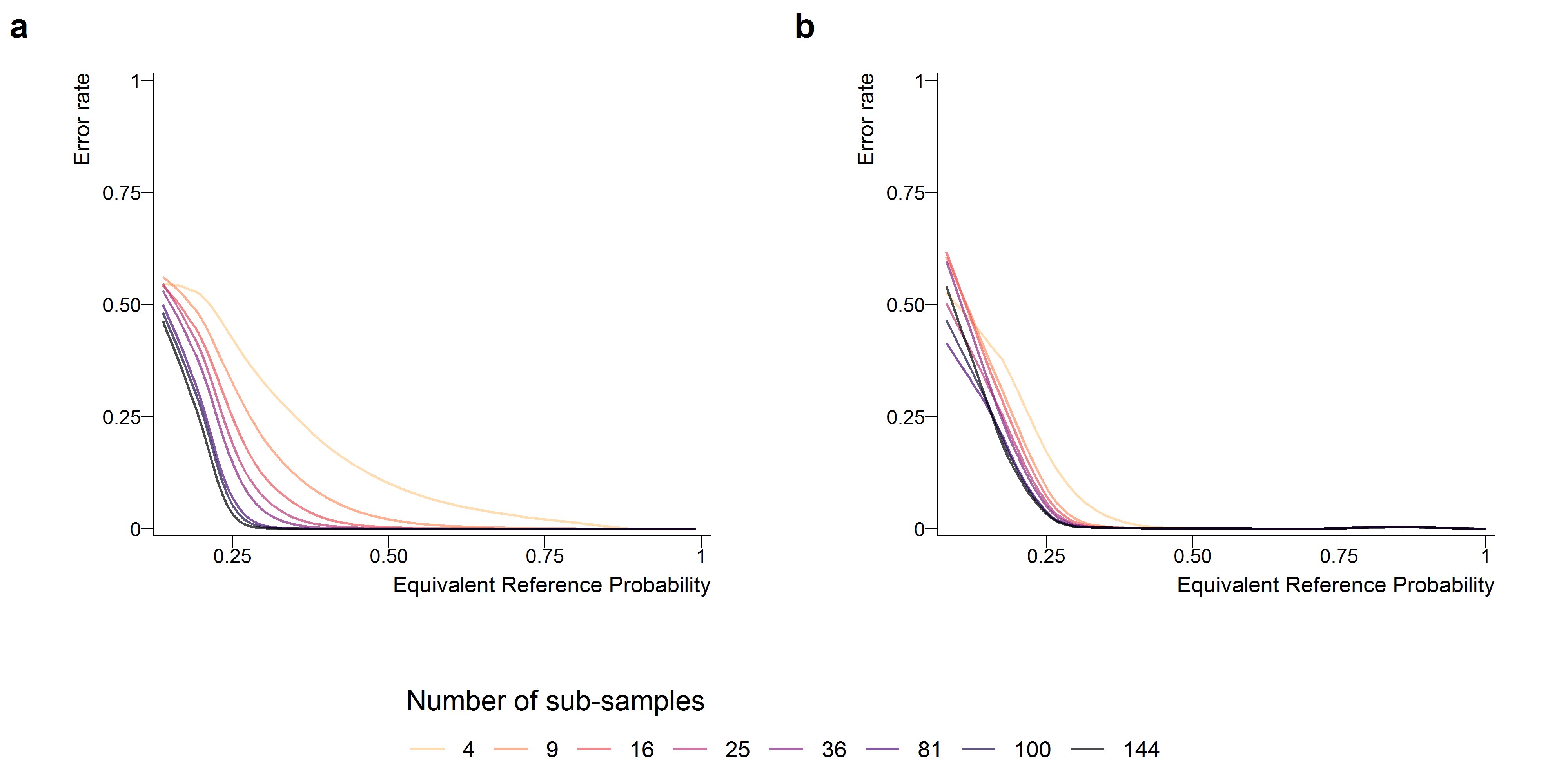}
\caption{\label{fig:error_rate_erp} Smoothed error rates in reference labels as a function of landscape fragmentation for majority point-based (a) and partition-based (b) designs.}
\end{figure*}

Heterogeneity indices allow us to generalise the overall error rates estimated on the study area. The selected heterogeneity indices, which are independent of the landscape and spatial resolution, highlight the peaks of the labelling uncertainty and the sampling units where the label can be trusted. The error rates are strongly related to the heterogeneity indices of the sampling units for both binary ($\pi$, see Fig.~\ref{fig:error_rate_nsamples_bin}) and majority legends ($\epsilon$, see Fig.~\ref{fig:error_rate_erp}). \\

In point-based designs, the error rate is maximum for sampling units with forest proportions close to the class threshold. The error distribution is slightly asymmetric, especially with small sampling efforts, with the longest tail towards the proportion of $50\%$.  Consistently with the expression of the variance for a binomial distribution, the largest variance of the distribution of errors is observed with the 50\% threshold and decreased towards the extremities of the range. In order to achieve an error rate of $< 1\%$ on average with $16$ points, the actual proportion need to be different from the threshold value by circa $20\%$. \\

The error distribution of partition-based response designs also peaks near the threshold values. The results clearly indicate that the partition-based method are strongly biased when the threshold value is not $50\%$: with $10$ and $75\%$ thresholds, the errors rates increase from 50\% towards the value of the threshold, where they are systematically wrong. This is due to the systematic omission of the class that contains the $50\%$ interval when approaching the extremities of the range, which is the only type of error when using a partition-based response design with these threshold values. Indeed, the error rate drops to zero in terms of detection of the class that is located at one end of the interval.  \\

With majority legends, the largest error rates are observed for sampling units with similar proportions (low $\epsilon$ value). Labelling is $99\%$ correct when $\epsilon\geq 0.5$. Interestingly, point-based designs are more accurate than partition-based designs for complex landscapes. On the other hand, designs based on a limited number of partitions outperformed their point-based counterparts for simple landscapes ($\epsilon>$0.25). Overall, partition-based designs appear relatively insensitive to an increase above $9$ in the number of sub-samples, while the impact of the number of sub-samples is obvious in the case of point-based design. This corroborates the results of overall error rates in the case study (Fig. \ref{fig:error_rate_nsamples}).\\

\subsection{Optimisation of the number of sub-samples}

We optimised the number of sub-samples so that the resulting label reached a $90\%$ or $99.9\%$ confidence level for each sampling unit or the maximum number of sub-samples (144) was attained. The difference between the error rates of the $99.9\%$ optimised and the error rate with a fixed ($144$) number of sub-samples was $<$1\%.\\

\begin{figure*}[htb!!!]
\centering
\includegraphics[width=0.85\linewidth]{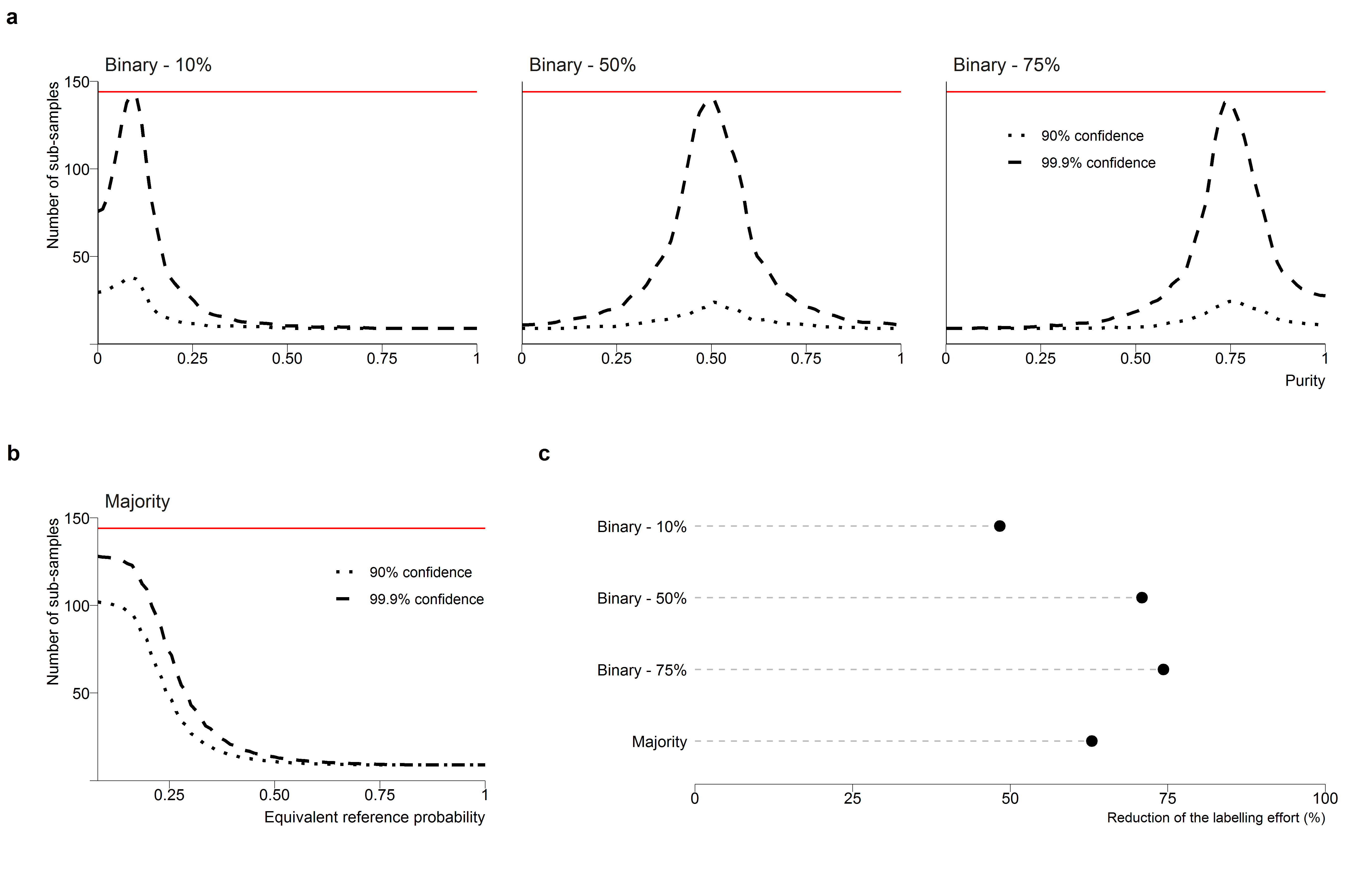}
\caption{\label{fig:benefit_optimisation} Effect of optimising the number of sub-samples on the labelling effort and the error rate using different confidence levels. The benchmark (no optimisation) is based on 144 sub-samples.}
\end{figure*}

The regions of label uncertainty highlighted in Fig. \ref{fig:error_rate_nsamples_bin} and Fig. \ref{fig:error_rate_nsamples} are consistent with the regions that require more validation efforts (Figure \ref{fig:benefit_optimisation}a and b). More samples are needed when $\epsilon$ is low or $\pi$ is close to the threshold value. The mean number of sub-samples then decreases quickly, especially with the binary legends, so that less than $20$ points are needed for most of the range of  $\epsilon$ or $\pi$ when the confidence level is set to $99.9\%$. Furthermore, the $90\%$ confidence level is achieved with low effort for the binary legends (Fig.~\ref{fig:benefit_optimisation}a). The sampling effort around the minimum $\epsilon$ value for the majority-based legend remains high in comparison (Fig.~\ref{fig:benefit_optimisation}b), which is consistent with the larger error rates observed for the point-based validation of majority-based legends. \\

The main difference between the shapes of the distribution of the error rates (Fig. \ref{fig:error_rate_nsamples_bin}) compared with the mean optimized number of points (Fig. \ref{fig:benefit_optimisation}a) occurs on the extreme values of $\pi$ for the binary legends. Indeed, the error rate for $\pi$ value of $0\%$ for the binary legend with $10\%$ threshold (or $100\%$ with threshold $75\%$) is close to zero, but the number of subsamples needed to achieve $99.9\%$ confidence is relatively high ($75$ for the $10\%$ threshold and $40$ for the $75\%$ threshold).\\

In the study area, the optimisation method with a very high confidence ($99.9\%$) could more than halve the labelling effort compared to a systematic sub-sampling of $144$ sub-samples per sampling unit (Fig.~\ref{fig:benefit_optimisation}c). It required an average of $57$, $27$ and $26$ sub-samples for binary thresholds of $10$, $50$ and $75\%$, respectively. The majority-rule legend was the most difficult to validate, with an average of $115$ points needed and the maximum ($144$) number of sub-samples needed in $56.1\%$ of the sampling units. \\

The $90\%$ confidence interval could be achieved for the binary legends with on average $21$, $10$ and $11$ sub-samples for thresholds of $10$, $50$ and $75\%$ respectively. With the majority-based legend, it required an average of $99$ points. Those results are due to the fact that the maximum number of sub-samples (that is $144$ in this study) was reached $38\%$ of the cases.  \\

\begin{figure*}[htb!]
\centering
\includegraphics[width=0.93\linewidth]{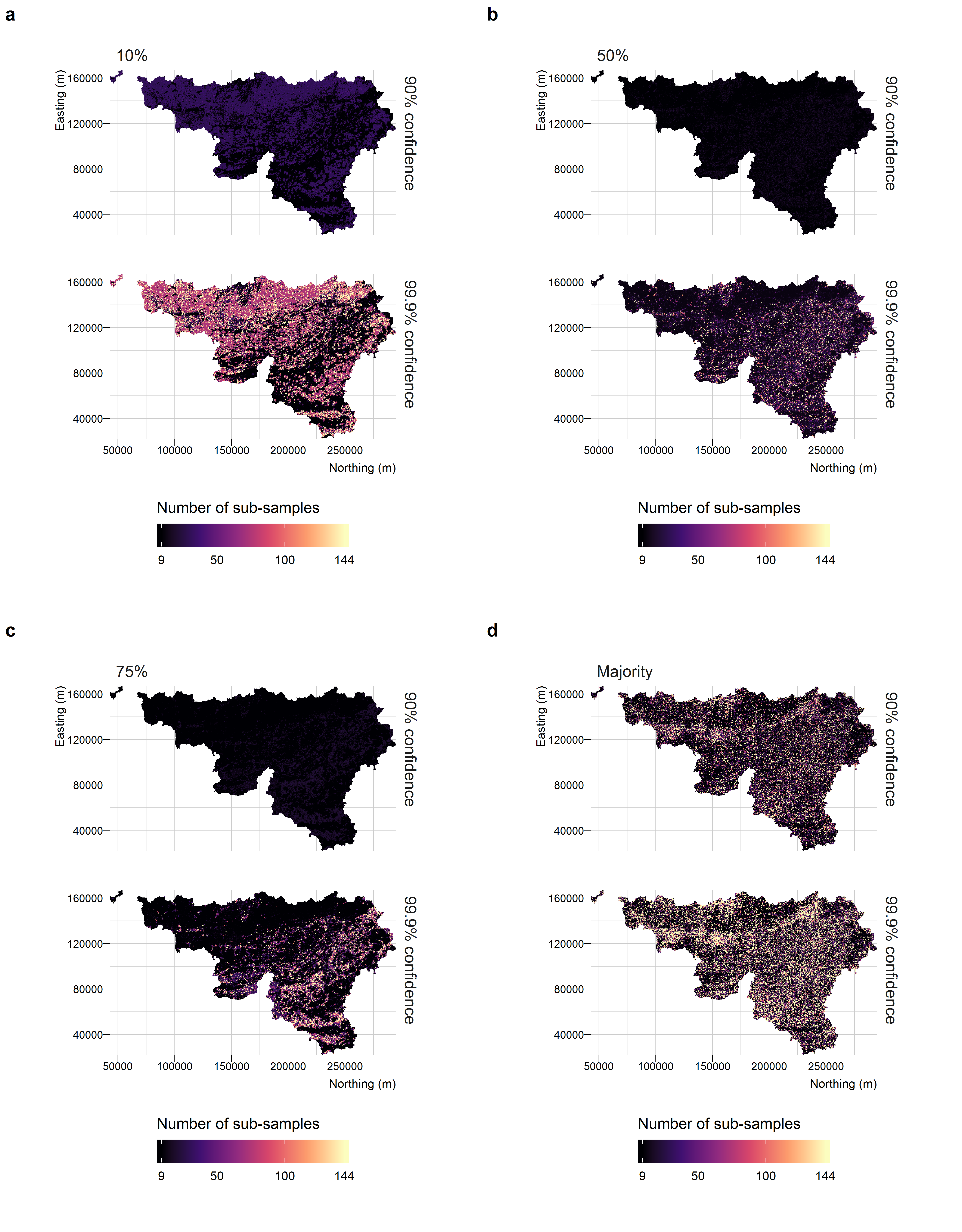} 
\caption{\label{fig:optimisation_pointbased} Optimised number of sub-samples for binary legends at 10\% (a), 50\% (b), 75\% (c), and for a majority legend (d). }
\end{figure*}

As observed in the Fig\ref{fig:optimisation_pointbased}, Fig.~\ref{fig:optimisation_pointbased} shows that reaching a confidence level of $90\%$ can be achieved at a reasonable cost, but the a large confidence level is much more demanding. However, the spatial distribution of the required number of subsamples (Fig.~\ref{fig:optimisation_pointbased}) highlights the particularities of the landscape in the study area. For instance, the majority-based legend requires more points in heterogeneous areas (such as the large urban areas) while the binary legends is more demanding in open landscapes (for the $10\%$ threshold) or closed forests (for the $75\%$ threshold), when the actual proportions are close to the threshold values. On the other hand, the benefits of the optimisation of the number of samples is obvious on large patches where the land cover proportions are distinct from the threshold values of the legend.  \\

\section{Discussions}
\label{sec:discussion}


We assessed the accuracy of two main types of quantitative response designs --a grid of points and a grid of squares-- based on a protocol that provided full control over the validation process. While it is well known that mixed pixels are more difficult to label than pure ones, we quantified how the labelling uncertainty increased for class proportions close to the decision boundaries of the legend. These results highlighted the difficulty of building accurate reference data sets for any combination of response design and legend. In many cases, the required number of sub-samples to reach $98\%$ confidence level was indeed too high (more than 100 subsamples per sampling unit) to be practically implemented. When factoring in the cost of response designs with large number of sub-samples, collecting error-free reference data seems thus barely feasible. Therefore, matching the data collection effort to the available resources appears critical. In other words, there is a necessary sacrifice of the confidence on the reference data in order to achieve a rigorous quality assessment at a reasonable cost. \\

The efficiency of point-based and partition-based response designs differed depending on the legend type. Partition-based response designs ought to be preferred in case of majority-based legend or binary legend when the proportion is close to 50\%, while point-based response designs become more efficient when the binary legends use thresholds that are close to $0$ or $100\%$. The ability to directly determine the class proportions inside a sampling could also help to arbitrate between the two types of partition-based response designs, MTT and TTM, because TTM is much more dependent of the operator skills than the other response designs. \\

The main advantage of point-based validation is the possibility to estimate the reliability of the label from the points themselves, and hence to objectively optimise the sub-sampling process without prior knowledge about the sampling units. We demonstrated that relying on a fixed number of sub-samples is inefficient because the same amount of resources is spent to label both obvious and complex sampling units. An efficient approach would reduce the effort for those easy-to-interpret cases and allocate it to label complex cases so as to increase their confidence. To this aim, we proposed to iteratively interpret sub-samples until the estimated class proportions reached the desired confidence level. Combined with advanced validation applications, such an approach computes the required number of sub-samples on the fly, thereby reducing the labelling cost in the case of obvious sampling units. We showed that, in our study area, the labelling effort could be reduced by 50\% to 75\% without affecting the accuracy of the labels. As a result, the labelling effort was strongly reduced across the study site and concentrated in the fragmented and ambiguous areas. In some cases, \ie close to the legend cut-off value, the added value of labelling additional points plateaus because sampling units with proportions close to the legend definition are always uncertain. It is therefore encouraged to set a threshold on the maximum labelling effort. \\

An iterative optimisation approach for partition-based designs is impractical because labels could be contradictory when changing scale. Therefore, optimising partition-based designs would rather depend on subjective operator decision about the proportions she/he estimates inside each sampling unit. Nonetheless, well trained operators could be granted the ability to select a number of partitions based on their impression of the complexity of the landscape. This method is likely to work well for threshold values near $50\%$ and could avoid extra work in simple cases, but remains sensitive to the Modifiable Area Unit Problem (MAUP) --a statistical biasing effect that occurs when arbitrary units are used to collect data such as class proportions. As described in \citet{jelinski1996modifiable}, the MAUP applies to two types of problems which are relevant in partition-based designs. The first aspect of the MAUP is the ``scale problem'', where the same set of areal data is aggregated into several sets of larger areal units, with each combination leading to different data values. The second is the ``zoning problem'', where a given set of areal units is recombined into zones that are of the same size but located differently, again resulting in variation in data values. MAUP could be mitigated by generating partitions that correspond to actual image objects derived via segmentation \citep{waldner2019conflation}. Image segmentation is mainly justified in landscapes that can be divided in a small number of homogeneous patches, not in areas that are very fragmented at a larger scale than the sampling unit. However, one may loose control over the number of sub-samples generated by the segmentation algorithm, leading to unpractical labelling effort. Besides, the effect of delineation errors is difficult to predict, so that the accuracy of the response design should then be assessed with external data or again rely on an estimation provided by the operator.  \\

In this study, the reported error rates resulted only from imprecise estimation of the class proportions. There are, however, additional errors that should still be considered for a complete understanding of the response design reliability: the simplification of the pixel model which is a simplified representation of the area observed by remote sensing \citep{foody1996incorporating, hsieh2001effect, radoux2016sentinel,fisher1997pixel},  geolocation errors, and photo-interpretation errors. While we assumed that operators made no errors throughout the paper, their performance is in reality imperfect~\citep{powell2004sources, vancutsem2012harmonizing, see2013comparing, waldner2019conflation}. For instance, \citet{powell2004sources} concluded that five interpreters were required to agree upon a specific class. Human factors are responsible for no less than $20\%$ of the inter-individual differences in operator performance~\citep{van2014variability}. To be more realistic, errors rates should account for errors of interpretation of the landscape and, in partition-based designs, errors in estimating the area of each class. As such, the error rates reported in this study are thus lower bounds.\\

\begin{figure}[htb]
\centering
\includegraphics[width=0.9\linewidth]{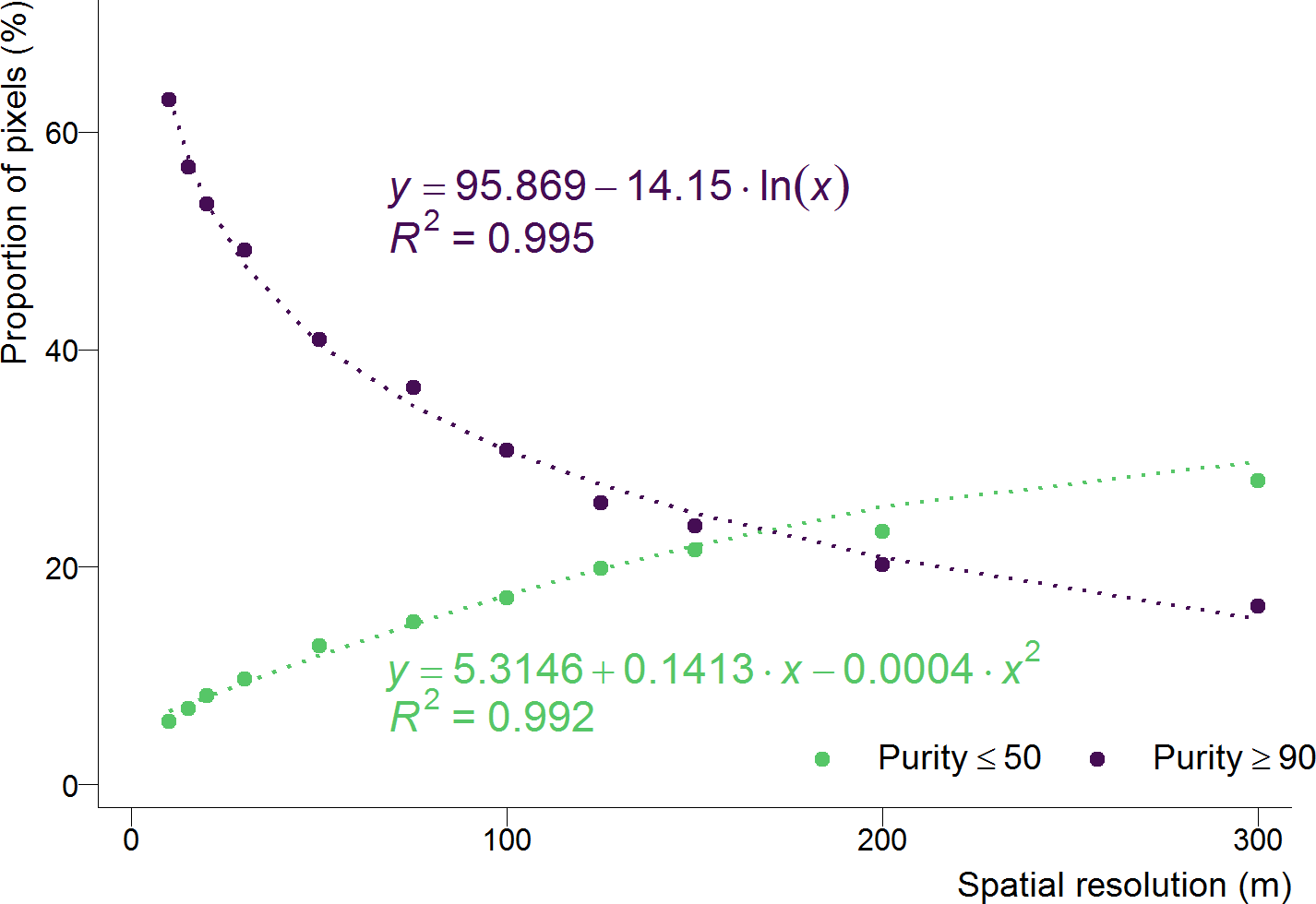} \caption{\label{fig:scalogram} Proportions of the purity index values for different spatial resolutions. The yellow line indicates pixels with a purity above 90\% and the blue line shows pixels with a purity of less than 50\%}
\end{figure}

Sampling units of 360 m $\times$ 360 m were used  because these are divisible by a large number of integers and, therefore, allowed us to easily simulate a large set of regular partition-based designs. While this practical constraint has no direct impact on the generalisation of the results, changing the size of sampling units would, however, indirectly impact the response design accuracy. Indeed, the average purity of the sampling units increases when the ratio between the ground sampling distance and the width of the object increases \citep{hsieh2001effect}. This general rule was also observed in our study area, which showed very strong relationship ($R^2>0.99$) between the pixel purity and the spatial resolution (Fig.~\ref{fig:scalogram}). If the legend remains the same, the accuracy of the response design will likely increase for sampling units of higher spatial resolution, and the estimation errors could be neglected when the sampling units become smaller than the spatial objects of interest. \\

In consolidating good practices to collect gold-standard validation data, we demonstrated that the amount of sub-samples required to meet stringent confidence levels is often too large to be realistically implemented. Therefore, this work suggests three main directions for future research. First, direct class assessment by operators should be compared with sub-sampling approaches to evaluate the overall level of confidence with real photo-interpretation in both cases. Second, unbiased confusion matrices could be built to account for uncertain reference data. While the errors affecting reference data cannot be predicted, we have however shown that the probability of error could be estimated based on the sub-samples. This information could be used to quantify a large part of the uncertainty of a reference data set at no extra cost. Third, the recent advances in image recognition and computer vision suggest that computer-assisted labelling of sub-samples could help to increase the number of sub-samples at lower cost~\citep[see][for instance]{xing2018exploring}. However, algorithms would need to perform at a high level of accuracy to avoid compromising the quality standards of reference data, and the accuracy of the secondary classification algorithm should also be assessed in order to measure the reliability of this kind of response design. \\

\section{Conclusions}

Photo-interpreted reference data sets are generally assumed error-free but they are in reality affected by erroneous labelling due to inaccurate image interpretation, drops of vigilance and estimation errors. We argued that, contrarily to interpretation errors and drops of vigilance that could be prevented (using for instance repeated-labelling), estimation errors are intrinsic to the response design and cannot be avoided once the response design is defined. With the goal of improving  good practices in reference data collection, we empirically assessed the relationship between the accuracy of reference data and the type of response design for binary and majority legends. Our results highlighted the need of a dense sub-sampling to obtain error-free reference data set.  We further evidenced that estimation errors are strongly linked to landscape composition, labelling errors being more prevalent when the class proportions are close to the class definition threshold (binary legends)  or in areas with complex class compositions (majority legends). To leverage the relationship between landscape composition and labelling accuracy, we proposed to iteratively interpret sub-samples until the class proportions are estimated with the desired confidence level. By quantifying the confidence of photo-interpreted labels, this optimisation method provides an efficient trade-off between the accuracy of the reference data and the labelling cost.  Therefore, its uptake by the remote-sensing community will likely result in more reliable accuracy estimates and subsequently improved assessment of the usability of thematic maps. \\

\section*{Acknowledgement}

This project was supported by the F\'ed\'eration Wallonie-Bruxelles (Lifewatch project) and by the Digiscape Future Science Plateform funded by the CSIRO. 

\section*{Conflict of interest}

None

\section*{References}

\bibliographystyle{elsarticle-harv}
\bibliography{sample.bib}


\end{document}